\documentclass[11pt]{article}
\usepackage{epsfig}

\def\leurre{\noindent\leftskip0pt\small\baselineskip 10pt}

\newtheorem{lem}{Lemma}
\newtheorem{thm}{Theorem}

\newtheorem{fig}{Figure}

\setbox221=\hbox{\includegraphics{cercle_L20.eps}}

\def\encercle#1#2{\hbox{\raise-5pt\copy221\hskip#2#1}}

\begin{document}

\def\ligne#1{\hbox to \hsize{#1}}
\def\PlacerEn#1 #2 #3 {\rlap{\kern#1\raise#2\hbox{#3}}}
\font\bfxiv=cmbx12 at 12pt
\font\itxiv=cmti12 at 12pt
\font\bfxii=cmbx12
\font\itxii=cmti12
\font\ttxi=cmtt11
\font\rmxi=cmr11
\font\ttx=cmtt10
\font\ttix=cmtt9
\font\pc=cmcsc10
\font\itix=cmti9
\font\bfix=cmbx9
\font\rmix=cmr9 \font\mmix=cmmi9 \font\symix=cmsy9
\def\mathix{\textfont0=\rmix \textfont1=\mmix \textfont2=\symix}
\font\rmviii=cmr8 \font\mmviii=cmmi8 \font\symviii=cmsy8
\def\mathviii{\textfont0=\rmviii \textfont1=\mmviii \textfont2=\symviii}
\font\rmvii=cmr7
%\font\rmviii=cmr8
\font\rmv=cmr5
\def\encadre#1#2{%
\setbox100=\hbox{\kern#1{#2}\kern#1}
\dimen100=\ht100 \advance \dimen100 by #1
\dimen101=\dp100 \advance \dimen101 by #1
\setbox100=\hbox{\vrule height \dimen100 depth \dimen101\box100\vrule}
\setbox100=\vbox{\hrule\box100\hrule}
\advance \dimen100 by .4pt \ht100=\dimen100
\advance \dimen101 by .4pt \dp100=\dimen101
\box100
\relax
}
\pagenumbering{roman}
\title{About the domino problem in the hyperbolic plane,
a new solution: complement}
\vskip 7pt
\author{Maurice Margenstern,\\
Universit\'e Paul Verlaine $-$ Metz,\\
LITA, EA 3097, IUT de Metz,\\
\^Ile du Saulcy,\\
57045 METZ C\'edex, FRANCE,\\
{\it e-mail}: {\it margens@univ-metz.fr}
}

\maketitle 
\pagenumbering{arabic}

\def\Hii{\hbox{$I\!\!H^2$}}
\def\cqfd{\hbox{\kern 2pt\vrule height 6pt depth 2pt width 8pt\kern 1pt}}
\vskip 15pt
\begin{abstract}
In this paper, we complete the construction 
of paper \cite{mmarXiv2,mmnewtechund}. Together with the
proof contained in \cite{mmarXiv2,mmnewtechund}, this paper
definitely proves 
that the general problem of tiling the hyperbolic plane 
with {\it \`a la} Wang tiles is undecidable.
\end{abstract}
\vskip 15pt

\def\cqfd{\hbox{\kern 2pt\vrule height 6pt depth 2pt width 8pt\kern 1pt}}
\vskip 15pt

\section{Introduction}

   The question, whether it is possible to tile the plane with copies of a fixed
set of tiles was raised by Wang, \cite{wang} in the late 50's of the
previous century. Wang solved the {\it partial} problem which consists in
fixing an initial finite set of tiles: indeed, fixing one tile is enough 
to entail the undecidability of the problem. The general case, later 
called the {\bf general tiling problem} in this paper, 
$GTP$ in short, without condition, in particular with 
no fixed initial tile, was proved undecidable by Berger in 1966, 
\cite{berger}.
Both Wang's and Berger's proofs deal with the problem in the Euclidean plane. 
In 1971, Robinson found an alternative, simpler proof of the undecidability of
the general problem in the Euclidean plane, see \cite{robinson1}. In this 1971 
paper, he raises the question of the general problem for the hyperbolic plane. 
Seven years later, in 1978, he proved that in the hyperbolic plane, the partial
problem is undecidable, see \cite{robinson2}. Up to now, and as far 
as I know, $GTP$ remained open.

   In this paper, we complete the proof that $GTP$ is 
also undecidable in the case of the hyperbolic plane which is
given in \cite{mmarXiv2,mmnewtechund}. 

   In a first section, we sketchilly remember the construction
of \cite{mmarXiv2,mmnewtechund} and we very briefly remember
the reader the construction of the mantilla and its properties
already proved in \cite{mmarXiv1,mmtechund}.

   In the second section, we give the needed complement. This will
completely prove that:

\begin{thm}\label{undec}
\it The general problem of tiling the hyperbolic plane is undecidable.
\end{thm}

   Then, we conclude with remarks on further improvements and a 
few corollaries
which we already obtained from the theorem.

   In this section, first, we very briefly mention the construction 
of the mantilla, the basic frame in which
the different implementations performed by our construction take
place.

   In the next sub-section, we briefly remind the {\bf abstract
brackets} which is the key new tool of the general frame of
the proof. This one-dimensional construction is mentioned in
\cite{levin}, and it is at the basis of Berger's proof of $GTP$
for the Euclidean plane. Robinson's 
proof of $GTP$ for the Euclidean
plane is based on a two-dimensional adaptation of the 
one-dimensional argument. Paper \cite{levin} focuses at the
two-dimensional construction and gives a deep account on this
situation, especially from an algebraic point of view.

   Then , we briefly look at implementation of the one-dimensional 
construction in the Euclidean plane, lifting the intervals
of this model into triangles. Such a construction is called
an {\bf infinite model}. This implementation is
transported into the hyperbolic plane, infinitely many times.
This entails a kind of cutting of the construction which we analysed
in our study of the one-dimensional construction under the name
of {\bf semi-infinite model}. The final step of the construction
consists in indicating a way to {\bf synchronize} all these 
implementations in such a way that they appear as different cuts
of a single infinite model. The key property of these triangles
is that there are infinitely many of them for infinitely many heights. 

   The last point is to implement a grid in each of these domains.
It allows to implement a space time diagram of the same Turing machine,
as in the classical proofs of Berger and Robinson. The complement
does not change this part of the proof which remains what it is
in \cite{mmarXiv2,mmnewtechund}.

The reader is invited to look at the technical report, 
\cite{mmnewtechund} 
on which the
paper is based and which is available at the following address:
\vskip 2pt
\ligne{\hfill\ttix
http://www.lita.sciences.univ-metz.fr/\~{}margens/new\_hyp\_dominoes.ps.gzip,
\hfill}

\noindent
where full proofs can be found of what is indicated in this
section.

\subsection{The mantilla}
\def\reunion{\mathop{\cup}}

   Here, we consider the tessellation $\{7,3\}$ of the hyperbolic 
plane, which we call the {\bf ternary heptagrid}, simply
{\bf heptagrid}, for short, see \cite{ibkmacri,mmJCA}. It is generated
by the regular heptagon with vertex angle 
$\displaystyle{{2\pi}\over3}$ by reflections in its sides and, 
recursively, of the images in their sides.

\subsubsection{The flowers}

In the ternary heptagrid, a {\bf ball} of {\bf radius}~$n$ around a 
tile $T_0$ is the set of tiles which are within distance~$n$ 
from~$T_0$  which we call the {\bf centre} of the ball. 
The {\bf distance} of a tile~$T_0$ to another $T_1$ is the number of 
tiles constituting the shortest path of adjacent tiles 
between~$T_0$ and $T_1$. We call {\bf flower} a ball of
radius~1.

\setbox110=\hbox{\epsfig{file=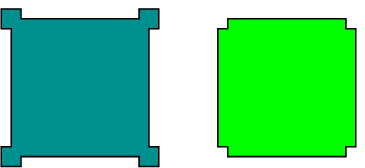,width=120pt}}
\setbox112=\hbox{\epsfig{file=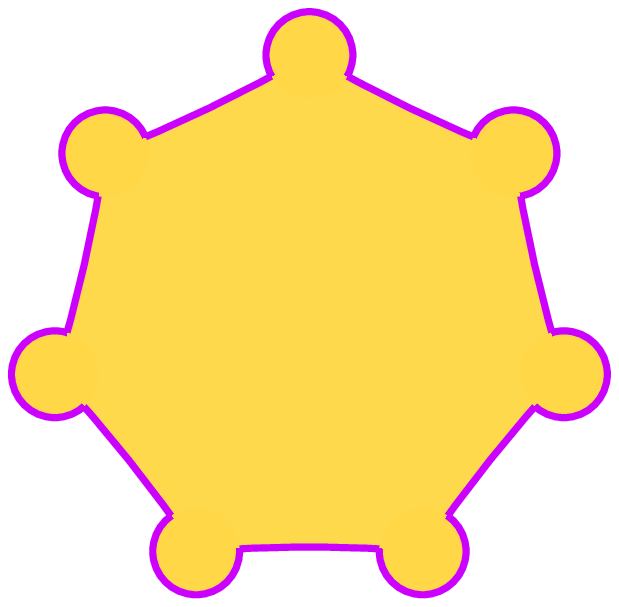,width=120pt}}
\setbox114=\hbox{\epsfig{file=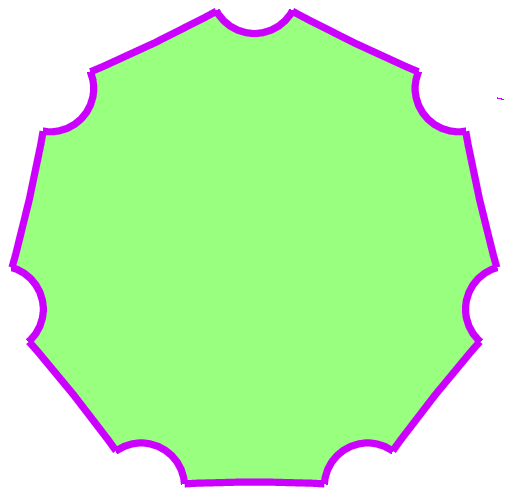,width=120pt}}
\ligne{\hfill
\PlacerEn {-350pt} {-80pt} \box110
\PlacerEn {-210pt} {-80pt} \box112
\PlacerEn {-120pt} {-70pt} {$a$}
\PlacerEn {-130pt} {-80pt} \box114
\PlacerEn {-40pt} {-70pt} {$b$}
}
\begin{fig}\label{robinsons}
\leurre
On the left: Robinson's basic tiles for the undecidability of the tiling 
problem in the Euclidean case. On the right: 
the tiles $a$ and $b$ are a 'literal' translation of Robinson's basic tiles to 
the situation of the
ternary heptagrid.
\end{fig}

\setbox110=\hbox{\epsfig{file=pseudo_robinson_a.ps,width=120pt}}
\setbox112=\hbox{\epsfig{file=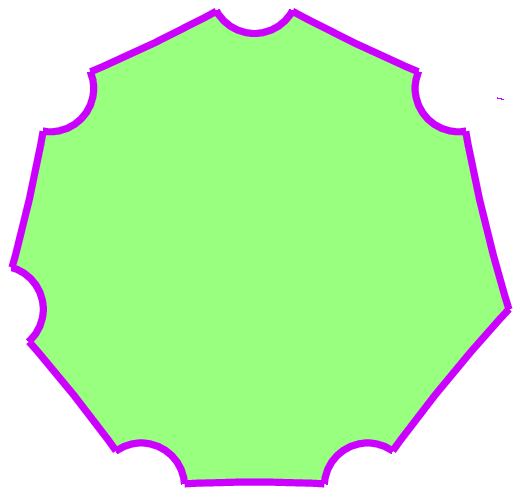,width=120pt}}
\setbox114=\hbox{\epsfig{file=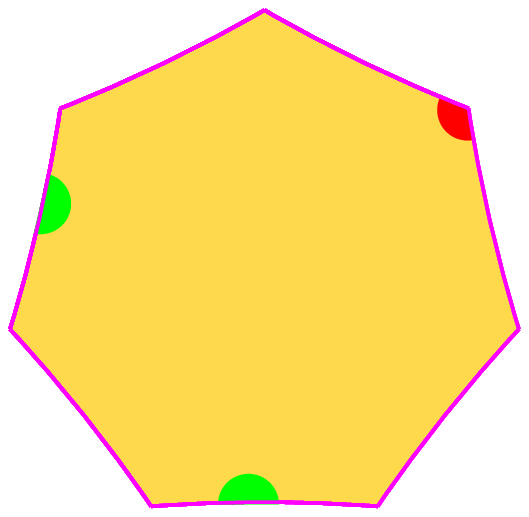,width=120pt}}
\setbox116=\hbox{\epsfig{file=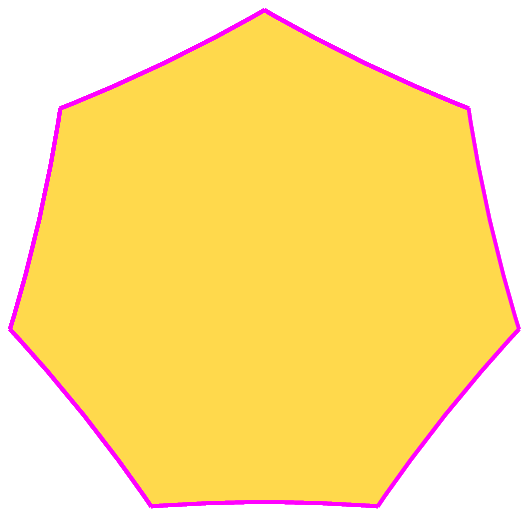,width=120pt}}
\ligne{\hfill
\PlacerEn {-385pt} {-80pt} \box110
\PlacerEn {-290pt} {-70pt} {$a$}
\PlacerEn {-295pt} {-80pt} \box112
\PlacerEn {-205pt} {-70pt} {$c$}
\PlacerEn {-205pt} {-80pt} \box116
\PlacerEn {-115pt} {-70pt} {$\alpha$}
\PlacerEn {-115pt} {-80pt} \box114
\PlacerEn {-25pt} {-70pt} {$\beta$}
}
\begin{fig}\label{from_rob_to_momo}
\leurre
On the left: change in the tiles {\it \`a la} Robinson. On the right:
their translation in pure Wang tiles.
\end{fig}

   The mantilla consists in merging flowers in a particular way. 
It comes from an attempt to implement Robinson's construction
in the Euclidean plane based on the left-hand side tiles of 
figure~\ref{robinsons}. The right-hand side tiles of the figure
are their 'literal' translation. It is not difficult to see that
it is not possible to tile the hyperbolic plane with tiles~$a$
and~$b$. However, a slight modification of the tile~$b$, see the 
tile~$c$ in figure~\ref{from_rob_to_momo}, leads to the solution.

   On the right hand side of figure~\ref{from_rob_to_momo}, we have 
the transformation of tiles~$a$ and~$c$ into Wang tiles. We call the 
tile~$\alpha$ a {\bf centre} and the tile~$\beta$ a {\bf petal}. 
We refer the reader to \cite{mmtechund,mmnewtechund} for the
numbering technique allowing to force the tiles~$\beta$ to be put
around tiles $\alpha$. Now, a petal belongs to three flowers at the
same time by the very definition of the implementation. From
this, there is a partial merging of the flowers.

   It is not difficult to see that there can be several types of 
flowers, considering the number of red vertices for which the other 
end of an edge is a vertex of a centre. We refer the reader to
\cite{mmtechund} for the corresponding properties. Here,
we simply take into consideration that we have three basic
patterns of flowers, which we call $F$-, $G$ and {\bf 8}-flowers
respectively. They are represented by figure~\ref{til_mantilla}.

   The figure also represents the way which allows to algorithmically
construct the tiling resulting from the tiles~$\alpha$
and~$\beta$ which we call the {\bf mantilla}. It consists in
splitting the {\bf sectors} generated by each kind of flowers
in sub-sectors of the same kind and only them, which we call the
{\bf sons} of the flower. From this, we easily
devise a way to recursively define a tiling. The construction is
deterministic below the flower, and it is non-deterministic 
when we proceed upwards. We do not make the notion of top and bottom
more precise: it will be done later. The exact description of
the splitting can be found in \cite{mmtechund}. We simply remark
that such a splitting is an application of the general method
described in \cite{mmDMTCS}, for instance.

\vskip 5pt
\setbox110=\hbox{\epsfig{file=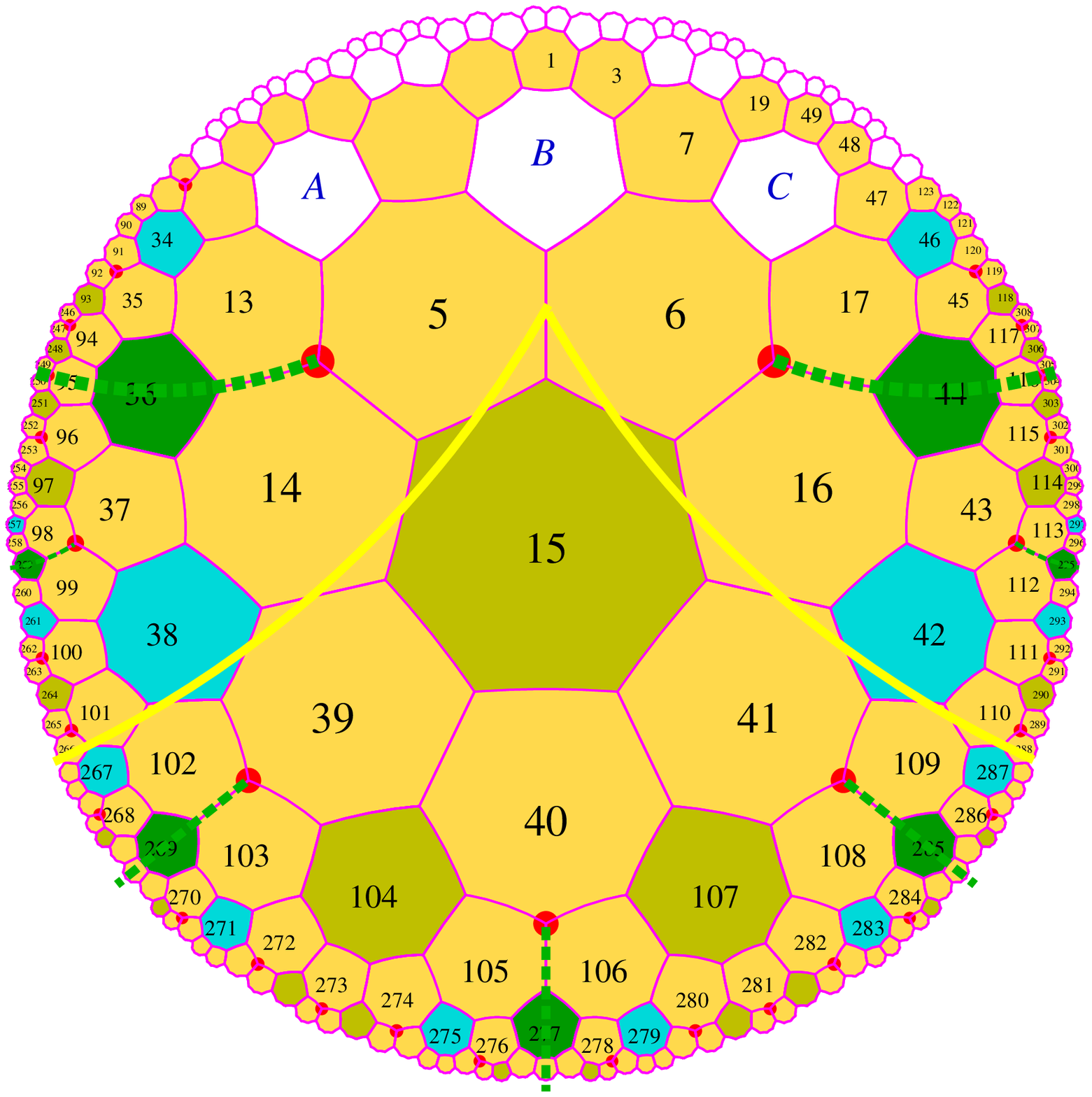,width=100pt}}
\setbox112=\hbox{\epsfig{file=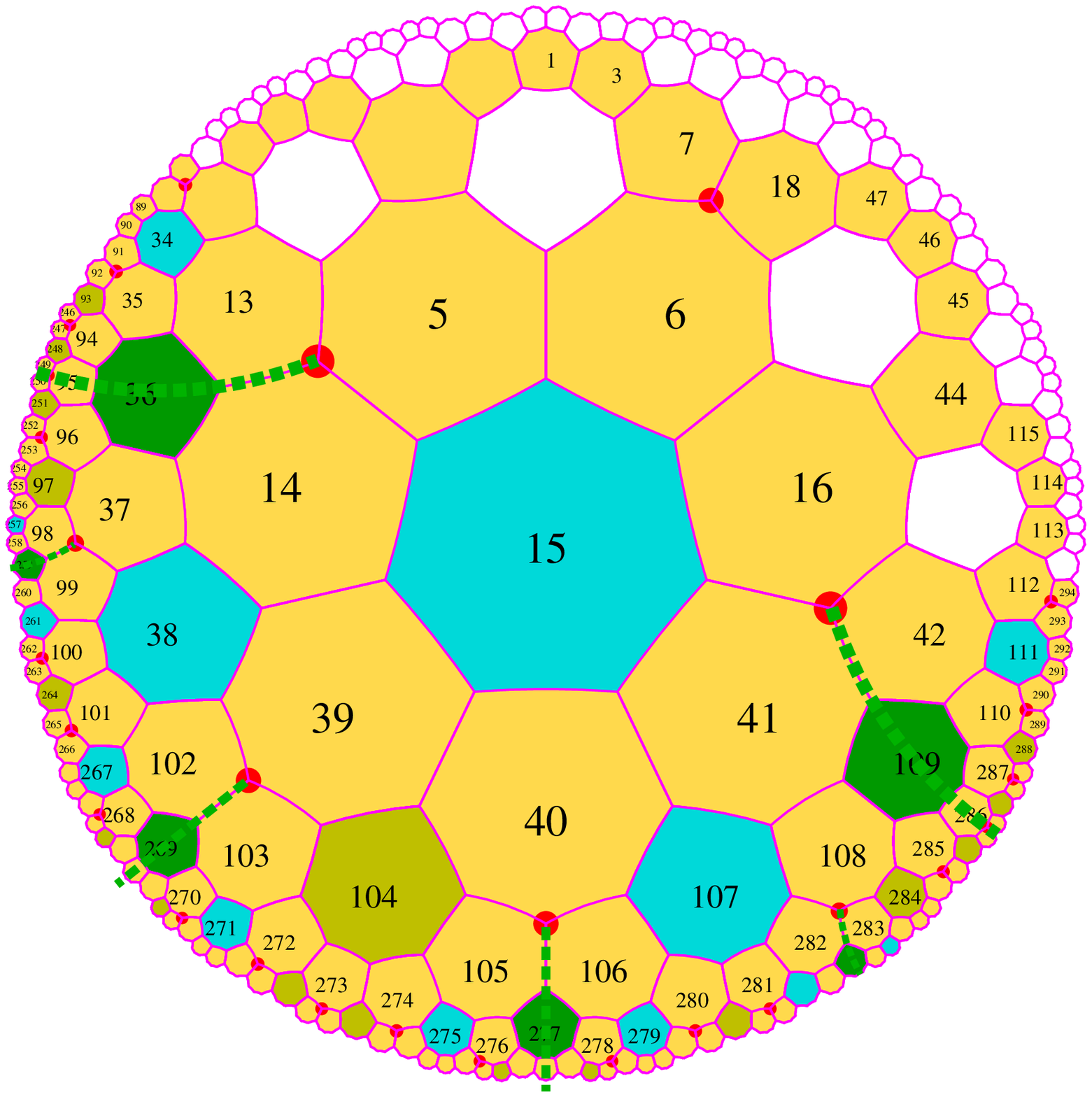,width=100pt}}
\setbox114=\hbox{\epsfig{file=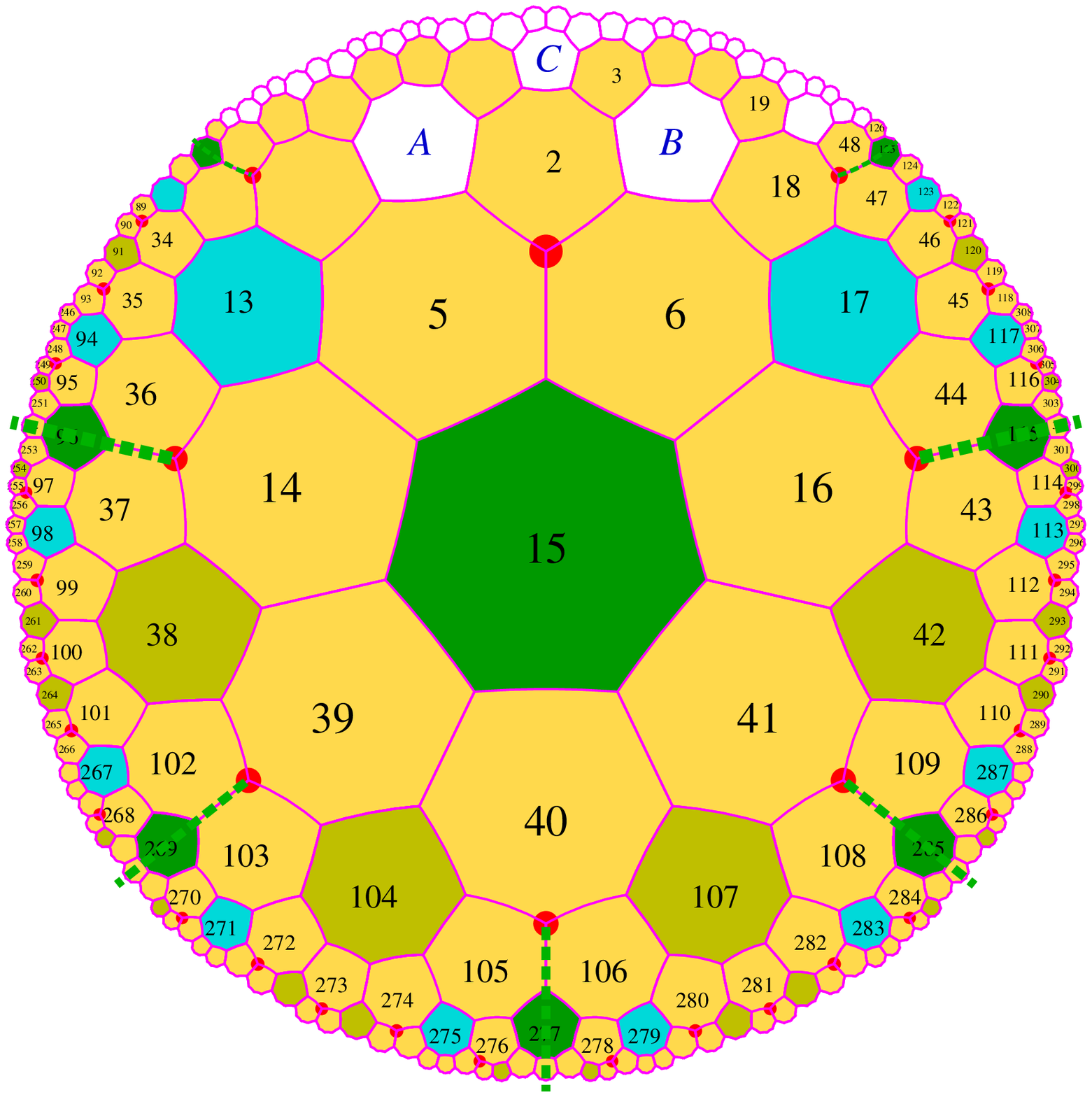,width=100pt}}
\ligne{\hfill
\PlacerEn {-365pt} {-60pt} \box110
\PlacerEn {-275pt} {-55pt} {$F$}
\PlacerEn {-245pt} {-60pt} \box112
\PlacerEn {-155pt} {-55pt} {$G$}
\PlacerEn {-125pt} {-60pt} \box114
\PlacerEn {-35pt} {-55pt} {\bf 8}
}
\begin{fig}\label{til_mantilla}
\leurre
Splitting of the sectors defined by the flowers. From left to right:
an $F$-sector, $G$-sector and {\bf 8}-sector.
\end{fig}

   Based on these considerations, we have the following result
which is thoroughly proved in \cite{mmtechund}:
\vskip 5pt

\begin{lem}
\label{skel_mantilla}
There is a set of $4$ tiles of type~$\alpha$ and $17$ tiles of
type~$\beta$ which allows to tile the hyperbolic plane as a mantilla. 
Moreover, there is an algorithm to perform such a construction.
\end{lem}

\subsubsection{Trees of the mantilla}

   Note that the left-most flower of figure~\ref{til_mantilla},
which represents an $F$-sector, also indicates a region delimited
by continuous lines, yellow in coloured figures. This lines 
are {\bf mid-point} lines, which pass through mid-points of 
consecutive edges of heptagons of the heptagrid. As shown in
\cite{ibkmacri,mmJCA}, they delimit a Fibonacci tree. The tiles
inside the tree which are cut by these mid-point rays are called
the {\bf borders} of the tree, while the set of tiles spanned
by the Fibonacci tree is called the {\bf area} of the tree.

   Say that an $F$-son of a $G$-flower is a {\bf seed} and the
tree, rooted at a seed is called a {\bf tree of the mantilla}. As the
seeds are the candidates for the construction of a computing region,
they play an important r\^ole. From figure~\ref{til_mantilla} we can
easily define the {\bf border} of a sector which is a ray crossing
{\bf 8}-centres. See \cite{mmtechund} for exact definitions.

\begin{lem}
\label{tree_sector}
The borders of a tree of the mantilla never meet the border of a
sector.  
\end{lem}

From 
lemma~\ref{tree_sector}, as shown in \cite{mmtechund}, 
we easily obtain:

\begin{lem}
\label{order_tree}
Consider two trees of the mantilla. Their borders never meet. 
Either
their areas are disjoint or the area of one contains the area of the
other.
\end{lem}

   From this, we can order the trees of the mantilla 
by inclusion of their areas.
It is clear that it is only a partial order. We are interested
by the maximal elements of this order. We call them {\bf threads},
see \cite{mmtechund} for an exact definition. Threads are indexed
by $I\!\!N$. Among them, there can be a unique {\bf ultra-thread}
which is indexed by $Z\!\!\!Z$. Note that the union of the areas of 
the trees which belong to an ultra-thread is the hyperbolic plane.
There can be realizations of the mantilla with or without an 
ultra-thread.

\subsubsection{Isoclines}

   In \cite{mmnewtechund}, we have a new ingredient. We define 
the status of a tile as {\bf black} or {\bf white}, defining
them by the usual rules of such nodes in a Fibonacci tree.
Then, we have the following property:

\begin{lem}
\label{black_tile}
If a seed is a black tile, all other seeds in the area of the tree
of the mantilla which it delimits are black tiles. Also, within
the same area, the {\bf 8}-centres are all black tiles.
\end{lem}
\vskip 5pt
\setbox110=\hbox{\epsfig{file=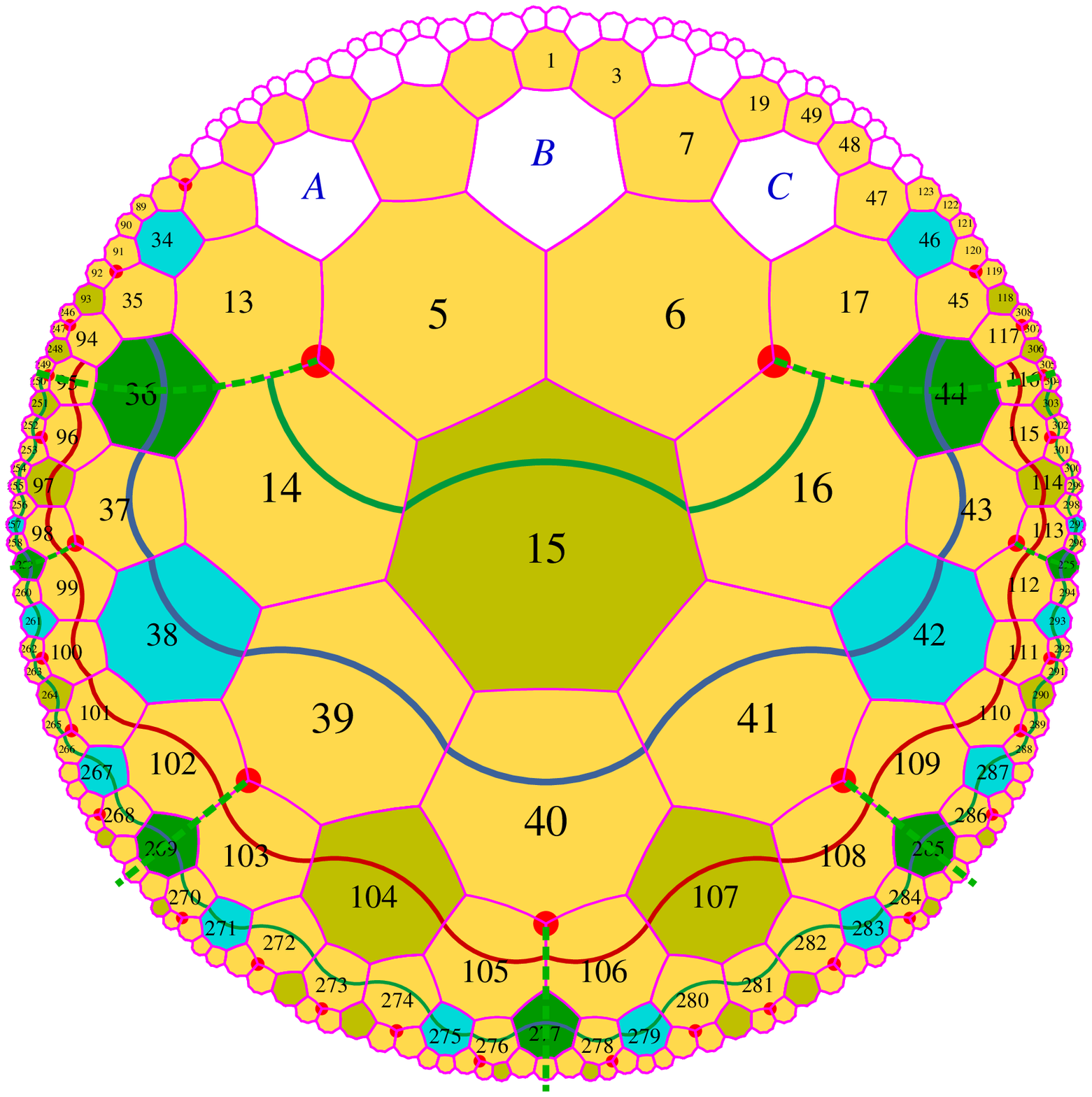,width=110pt}}
\setbox118=\hbox{\epsfig{file=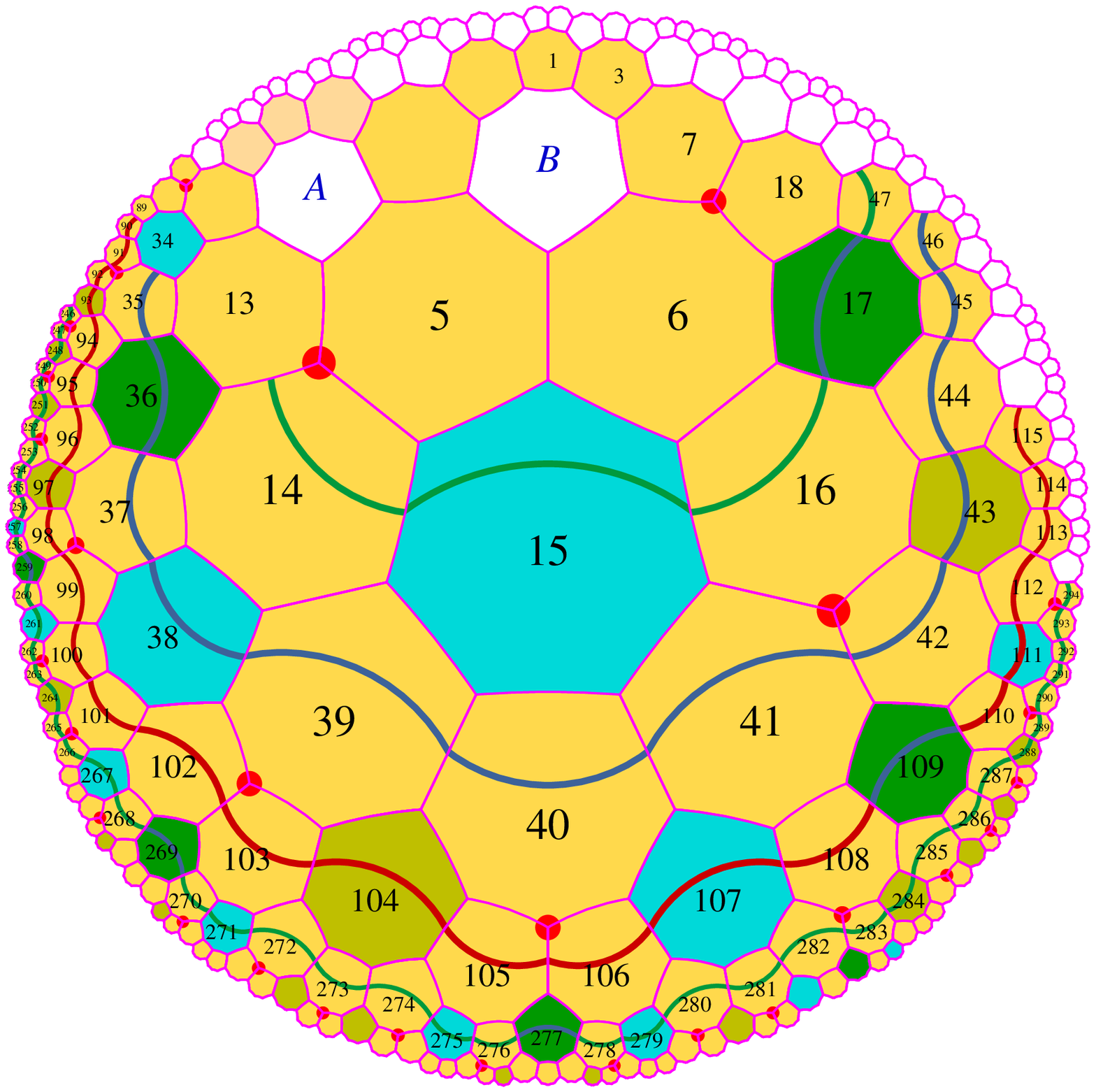,width=110pt}}
\ligne{\hfill
\PlacerEn {-340pt} {0pt} \box110
\PlacerEn {-180pt} {0pt} \box118
}
\begin{fig}\label{blackseedfig}
\leurre
The black tile property and the levels:
\vskip 0pt\noindent
On the left-hand side, a black $F$-centre; on the right-hand side,
a black $G_\ell$-centre. We can see the case of an {\bf 8}-centre on both
figures.
\vskip 2pt
\end{fig}

   As shown in \cite{mmnewtechund}, we can define arcs as follows:
in a white tile, the arc joins the mid-points of the sides which
have a common vertex with the side shared by the father. 
In a black tile, the arc joins the mid-point of the
sides which are separated by the side shared by the father and the
side shared by the uncle, which is on the left-hand side of the
father. 
Joining the arcs,
we get paths. The maximal paths are called {\bf isoclines}. 
They are illustrated in figure~\ref{blackseedfig}.
An isocline
is infinite and it splits the hyperbolic plane into two infinite 
parts.  
The isoclines from the different trees match, even when the areas are
disjoint.

\begin{lem}
\label{iso5}
Let the root of a tree of the mantilla~$T$ be on the isocline~$0$.
Then, there is a seed in the area of~$T$ on the isocline~$5$. 
If an {\bf 8}-centre~$A$ is on the isocline~$0$, starting from the
isocline~$4$, there are seeds on all the levels. From the 
isocline~$10$ there are seeds at a distance at most~$20$ from~$A$.
\end{lem}

   We number the isocline from 0 to 19 and repeat this, periodically.
This allows to give sense to {\bf upwards} and {\bf downwards} in
the hyperbolic plane.

\subsection{The abstract brackets}

   We refer the reader to \cite{mmnewtechund} for an exact definition.
However, figure~\ref{silentfig}, below, illustrates the construction
which now, we sketchily describe.

   The generation~0 consists of points on a line which are regularly
spaced. The points are labelled $R$, $M$, $B$, $M$, in this order,
and the labelling is periodically repeated. An interval defined by
an $R$ and the next~$B$, on its right-hand side, is called 
{\bf active} and an interval defined by a~$B$ and the next~$R$
on its right-hand side is called {\bf silent}. The generation~0 is
said to be {\bf blue}.

\vskip 5pt
\setbox110=\hbox{\epsfig{file=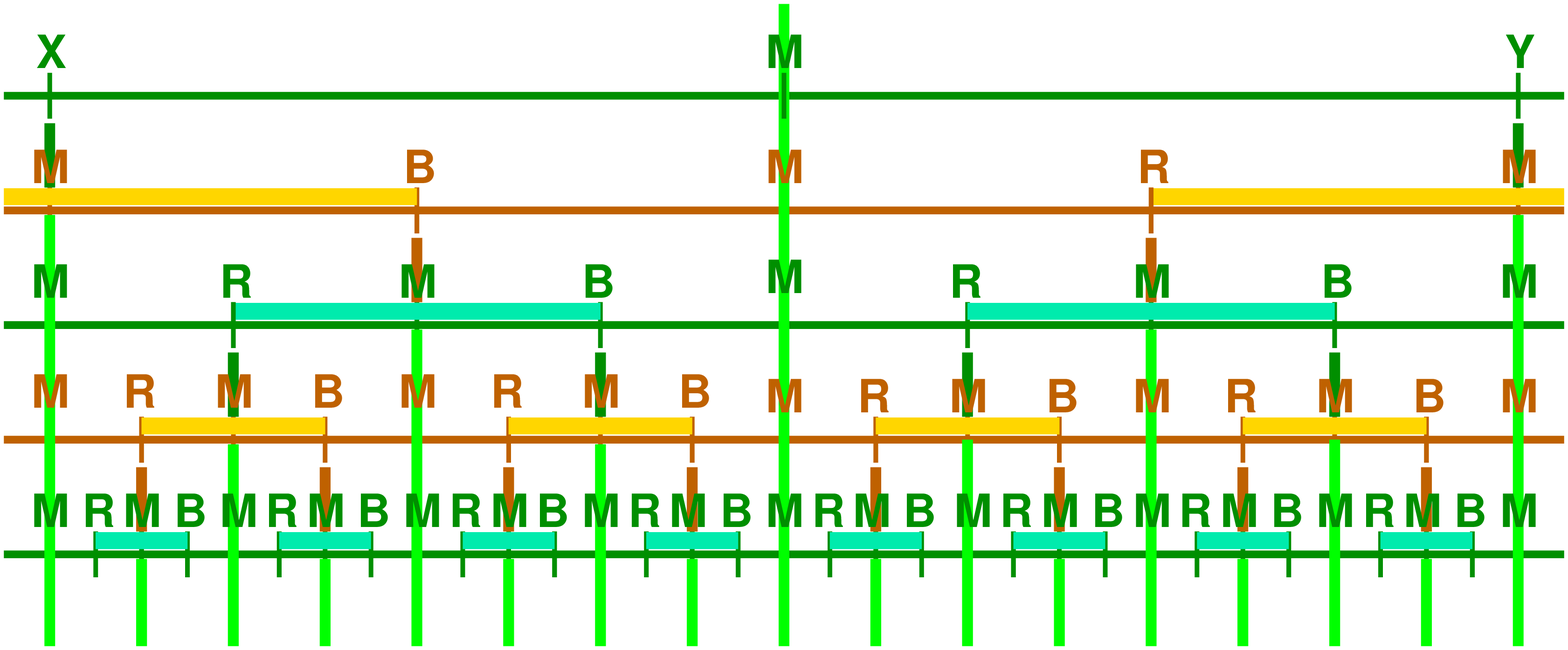,width=300pt}}
\ligne{\hfill
\PlacerEn {-327pt} {-10pt} \box110
}
\begin{fig}\label{silentfig}
\leurre
The silent and active intervals with respect to mid-point lines. The light green
vertical signals send the mid-point of the concerned interval to the next
generation. The colours are chosen to be easily replaced by red or blue inan opposite way. The ends $X$ and $Y$ indicate that the figure can be used
to study both active and silent intervals.
\end{fig}
\vskip 5pt
   Blue and red are said {\bf opposite}. Assume that the 
generation~$n$ is defined. For the generation~$n$+1, 
the points which we take into consideration are the points which are
still labelled~$M$ when the generation~$n$ is completed. Then, we
take at random an $M$ which is the mid-point of an active interval
of the generation~$n$, and we label it, either $R$ or~$B$. Next,
we define the active and silent intervals in the same way as for the 
generation~0. The active and silent intervals of the generation~$n$+1
have a colour, opposite to that of the generation~$n$.

   When the process is achieved, we get an {\bf infinite model}.
The model has interesting properties, see \cite{mmnewtechund}. 
We cannot mention all of them
here. We postpone some of them to the Euclidean implementation with
triangles. 

   In an interval of the generation~$n$, consider that a letter of
a generation~$m$, $m\leq n$, which is inside an active interval is 
hidden for the generations~$k$, $k\geq n$+1. Also, a letter has
the colour of its generation. Now, we can prove that in the blue
active intervals, we can see only one red letter, which is the 
mid-point of the interval. However, in a red active interval of the
generation~$2n$+1, we can see $2^{n+1}$+1 blue letters.

   Cut an infinite model at some letter and remove all
active intervals which contain this letter. What remains on the 
right-hand side of the letter is called a {\bf semi-infinite model}.

   It can be proved that in a semi-infinite model, any letter~$y$
is contained in at most finitely many active intervals, see 
\cite{mmnewtechund}.
      
\subsection{Interwoven triangles}

   Now, we lift up the active intervals as {\bf triangles} in the
Euclidean plane. The triangles are isoceles and their heights are
supported by the same line, called the {\bf axis}, see 
figure~\ref{interwoven_fig}.

   We also lift up silent intervals of the infinite model up to
again isoceles triangles with their heights on the axis. To distinguish
them from the others, we call them {\bf phantoms}. We shall speak of
{\bf trilaterals} for properties shared by both triangles and
phantoms.

\setbox110=\hbox{\epsfig{file=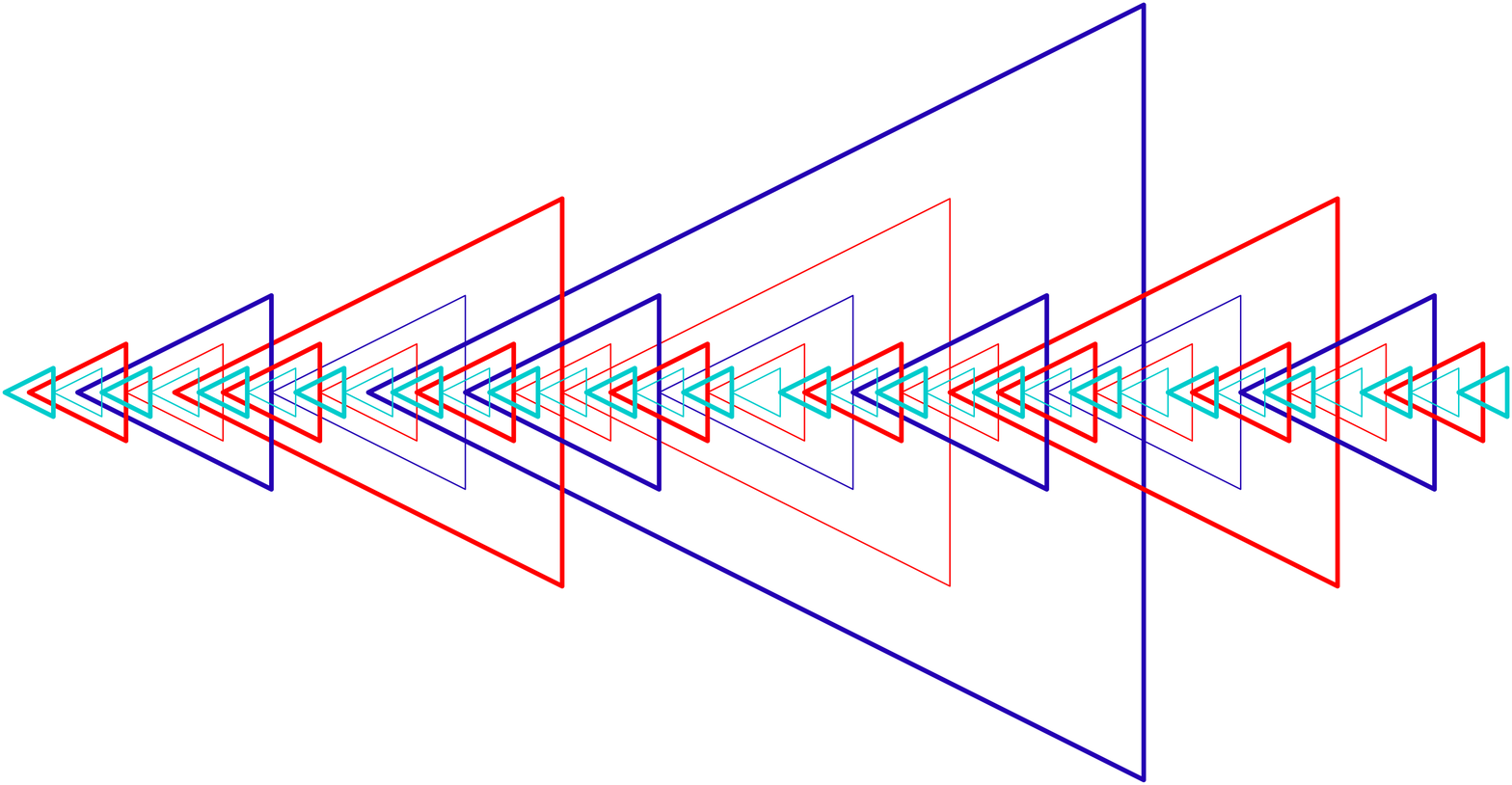,width=300pt}}

\ligne{\hfill
\PlacerEn {-350pt} {0pt} \box110
}
\vspace{-10pt}
\begin{fig}\label{interwoven_fig}
An illustration for the interwoven triangles.
\end{fig}

   We have very interesting properties for our purpose.

\begin{lem}\label{interwoven} 
Triangles of the same colour do not meet nor overlap: they 
are disjoint
or embedded. Phantoms can be split into {\bf towers} of 
embedded phantoms with the same mid-point and alternating colours.
Trilaterals can meet by a basis cutting the half of leg which
contains the vertex.  
\end{lem}

   From these properties, we prove in \cite{mmnewtechund} that:

\begin{lem}\label{tile_interwoven}
the Euclidean plane which can be forced by set of $190$ tiles.
\end{lem}

   In \cite{mmnewtechund}, we display the corresponding tiles which
are in a square format, and we also describe them with the help
of formulas taking into account the properties of
lemma~\ref{interwoven}.

\subsection{Hyperbolic implementation and the computing areas}

   We implement the interwoven triangles in the hyperbolic plane
by using the trees of the mantilla as frames for the legs of
the trilaterals. The basis is materialized by the trace of an
isocline in the area of the trilateral.

\subsubsection{The synchronization}

    The axis will be somehow materialized by a thread.
As most threads are indexed by $I\!\!N$ only, we have always
the implementation of a semi-infinite model. Now, we shall manage
the implementation in such a way that the semi-infinite models
are simply different cuts of the same infinite model. The possibility
of the realization of the infinite model in the case of an 
ultra-thread brings in no harm.

   To achieve this point, we very briefly indicate
a feature of the tiles. The legs of a triangle emit horizontal
signals {\bf outside} the triangle. The signals have the same
colour as the emitting triangle and they have a laterality. The 
left-hand side leg emits left-hand side signals, the right-hand side 
leg emits right-hand side ones. Both kinds of signal cross the
tiles in an {\bf upper} or {\bf lower} position, always at the lower
one for the vertex. Phantoms also emit signals,
only at the vertex, in a {\bf lower} position, and at the corner of 
the basis, in an {\bf upper} position.

   The tiling forces the construction of trilaterals generation 
after generation. A vertex of the next generation grows legs downwards
until they meet the green signal which indicates the mid-point of
the legs. Triangles stop their green signal, phantoms do not.  

   To synchronize the semi-infinite models, bases of triangles
which are on the same isocline merge. The distinction between 
outside and inside a triangle is given by the presence or absence
of the upper horizontal signal of the same colour as the basis.
We say that the basis is {\bf covered} or {\bf open}.
Inside a triangle, the left-hand side and right-hand side signals
can be joined only at a vertex, and so, they must be lower signals.
Outside a triangle, horizontal signals of different lateralities
can be joined, as the directions from where the signals come are
the opposite with respect to what happens inside a triangle. The 
needed tiles are provided only for meetings outside trilaterals.

   Now, the distinction between a covered and an open basis allows
the implementation of the construction using the tiles devised
for lemma~\ref{tile_interwoven}, using the same algorithm of
construction. Indeed, first halves of legs, {\it i.e.} from the
vertex to the mid-point, may cut bases, either covered or open,
leaving them covered or open respectively. The change to the 
second half is triggered
by the detection of the green signal. Next, the second half meets
covered bases. The first open basis, necessarily of its colour, is
the expected basis for this trilateral.

   Note that inside a trilateral and between the same set of isoclines,
there are several triangles of former generations. In the next
section we manage this point for which the complement given in 
this paper is needed.

\subsubsection{The computing areas}

   They are defined by the {\bf active} seeds which we now define.

   By definition, we decide that all seeds which are on an 
isocline~0 are active. This is enough to guarantee that the set of
active seeds is dense in the hyperbolic plane. Next, an active
seed diffuses a {\bf scent} inside its trilateral until the
isocline~5, starting from this seed, is reached. Seeds which receive
the scent, and only them, become active. An active seed also triggers
the green signal if and only if it reaches an isocline~5 or~15. 
By construction, The generation~0 is not determined
by the meeting of a green signal. But the others are. 

   We can see that the scent process constructs a tree. The branches 
of the tree materialize the thread which implements the considered
semi-infinite model. Note that the above synchronization mechanism
fixes things for spaces between triangles but also inside them.

   An important mechanism provided by the tiles of 
lemma~\ref{tile_interwoven} is the detection of the {\bf free
rows} inside {\bf red} triangles. These free rows are the isocline 
whose projection on the axis is a blue letter, visible in the active
interval defined by the height of the triangle. It is not difficult
to provide tiles for that, also based on the red horizontal signals
of different lateralities and positions, see \cite{mmnewtechund}.
      
\vskip 5pt
\setbox110=\hbox{\epsfig{file=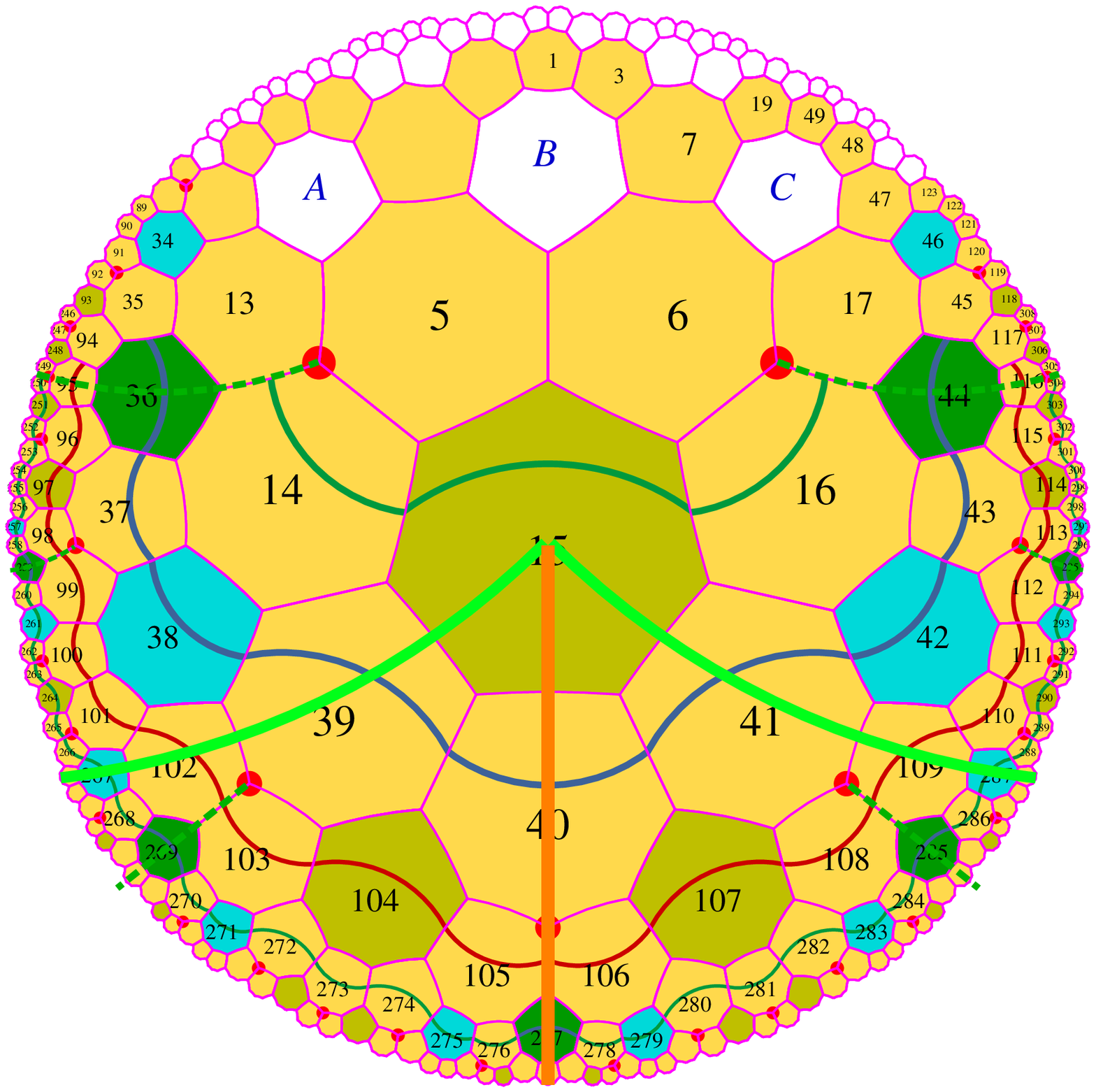,width=110pt}}
\setbox112=\hbox{\epsfig{file=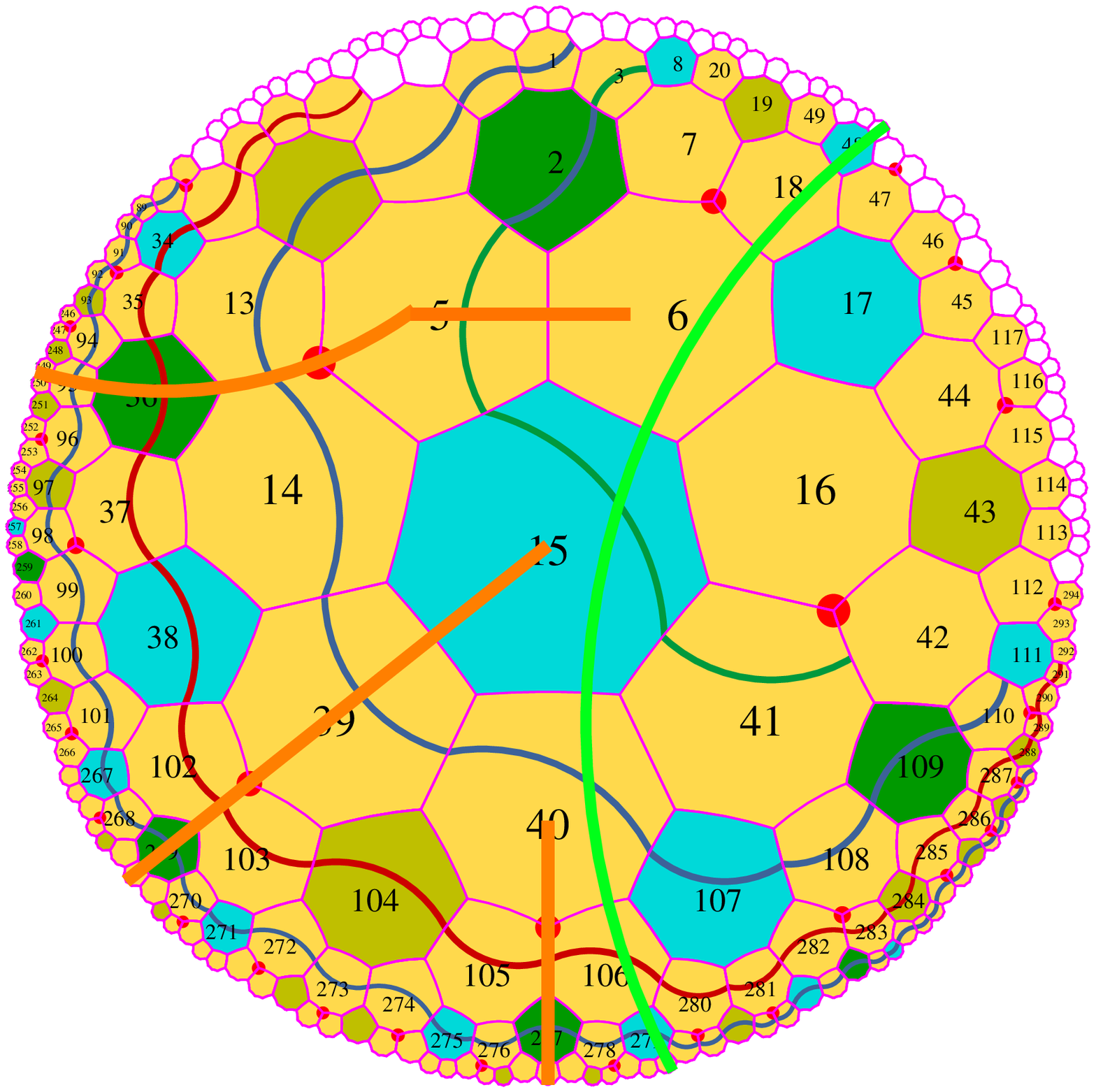,width=110pt}}
\ligne{\hfill
\PlacerEn {-130pt} {0pt} \box110
\PlacerEn {20pt} {0pt} \box112
\hfill}
\vskip-15pt
\begin{fig}\label{vertical_1}
\leurre
The perpendicular starting from a point of the border of a triangle which
represents a square of the Turing tape.
\vskip 0pt
On the left-hand side: the case of the vertex. On the right-hand side,
the three other cases for the right-hand side border are displayed on the
same figure.
\end{fig}

   The free rows inside the red triangles are the horizontal of our
computing areas. It remains us to define the verticals which are
needed for implementing the space-time diagram.

   The verticals consist into rays which cross {\bf 8}-centres.
Figure~\ref{vertical_1} illustrates how they are connected to
the different possible cases of contact of the isocline of a free
row with the border of the tree.

   The computing signal starts from the seed. It travels on
the free rows. Each time a vertical is met, which contains a symbol
of the tape, the required instruction is performed. If the direction
is not changed and the corresponding border is not met, the signal
goes on on the same row. Otherwise, it goes down along the vertical
until it meets the next free row. There, it looks at the expected
vertical, going in the appropriate direction. Further details
are dealt with in \cite{mmnewtechund} and are rather close to
the classical proofs.

\def\kvs{\vspace{-5pt}}

\section{The complement}

   In this section, we deal with the tuning promised in
sub-section~1.4.1, about the description of the
synchronization mechanism.

\subsection{The point to be tuned}

   In fact, in sub-section~1.4.1, we describe the synchronization
problems raised by the bases and vertices, in order to obtain
that all the threads implement a cut of the same infinite model
of the abstract brackets. Of course, if the bases and vertices
are synchronized, the mid-lines of the trilaterals are also
synchronized. As mentioned in sub-section~1.4.1, this time
we have infinitely many copies of the same trilateral within
a same set of isoclines: call such a set a {\bf latitude}. 
Now, we have to closer look at the possible interactions between 
the trilaterals which occur within a fixed latitude. Also
call {\bf amplitude} the number of isoclines contained in a
given latitude. The
problem is that in between two contiguous triangles of the
same latitude~$\Lambda$, there may be and, usually, there are 
trilaterals of smaller generations, whose latitude is contained 
in~$\Lambda$. Now, a part of the phantoms within~$\Lambda$ have
the same mid-line as the triangles whose height is the amplitude
of~$\Lambda$. Now, when we consider the phantoms which are crossed
by this mid-line between two consecutive triangles~$T_1$ and~$T_2$, 
the green line which they emit runs along the mid-line. As the
legs of a phantom do not stop the green signal, nothing 
prevent them to meet the legs of~$T_1$ and~$T_2$. Now, they should 
not meet these legs as the legs of a triangle stop the green 
signal. 

\subsection{A possible mechanism}

   The mechanism is the following. Any triangle~$T$ of a given
latitude~$\Lambda$ whose height is the amplitude of~$\Lambda$,
stops the green signal which runs on its mid-line. Now, outside
its legs, on the mid-line, $T$ stretches out two antennas: 
a left-hand- and a right-hand side one. They r\^ole is
to detect, the anteanna sent by the next triangle of the
same latitude in the direction followed by the antenna.

   What we need is a characterization of the structure of the
trilaterals within a given latitude. This is provided by the
following lemma:

\begin{lem}\label{horizontal}
Let $T$ and~$T'$ be two consecutive triangles of the same 
generation~$n$
and within the same latitude~$\Lambda$ whose amplitude is the
common height of~$T$ and~$T'$. Let $A$ be the mid-point of the
right-hand side leg of~$T$ and let~$B$ be the mid-point of the
left-hand side leg of~$T'$. Then, there is a tile~$C$ and a tile~$D$
on the isocline passing through~$A$ and~$B$ such that the interval
$[C,D]$ is contained in the interval $[A,B]$ and: if a leg of
a trilateral of a generation~$m$, with $m>n$, crosses $]A,C[$,
$]D,B[$ respectively, 
then it is a right-hand side leg, a left-hand side leg respectively, 
of this trilateral. Moreover, $[C,D]$ is not crossed by any leg 
of a trilateral belonging to a generation~$m$, with $m>n$.
A leg ending with a corner within $[A,C]$ or $[D,B]$
is considered as a crossing leg. 
\end{lem}

\noindent
Proof.
Let~$I$ be the isocline which contains~$A$ and~$B$.
Between~$A$ and~$B$, $I$ crosses several legs of trilaterals.
It crosses both legs of a trilateral if and only if the
trilateral belongs to a smaller generation: it is contained 
in~$\Lambda$. In this case, the trilateral is a phantom.
It cannot be a triangle: smaller triangles have their projection 
within the projection of a half leg of~$T$ and so, they cannot 
meet~$I$. 

\vskip 10pt
\setbox110=\hbox{\epsfig{file=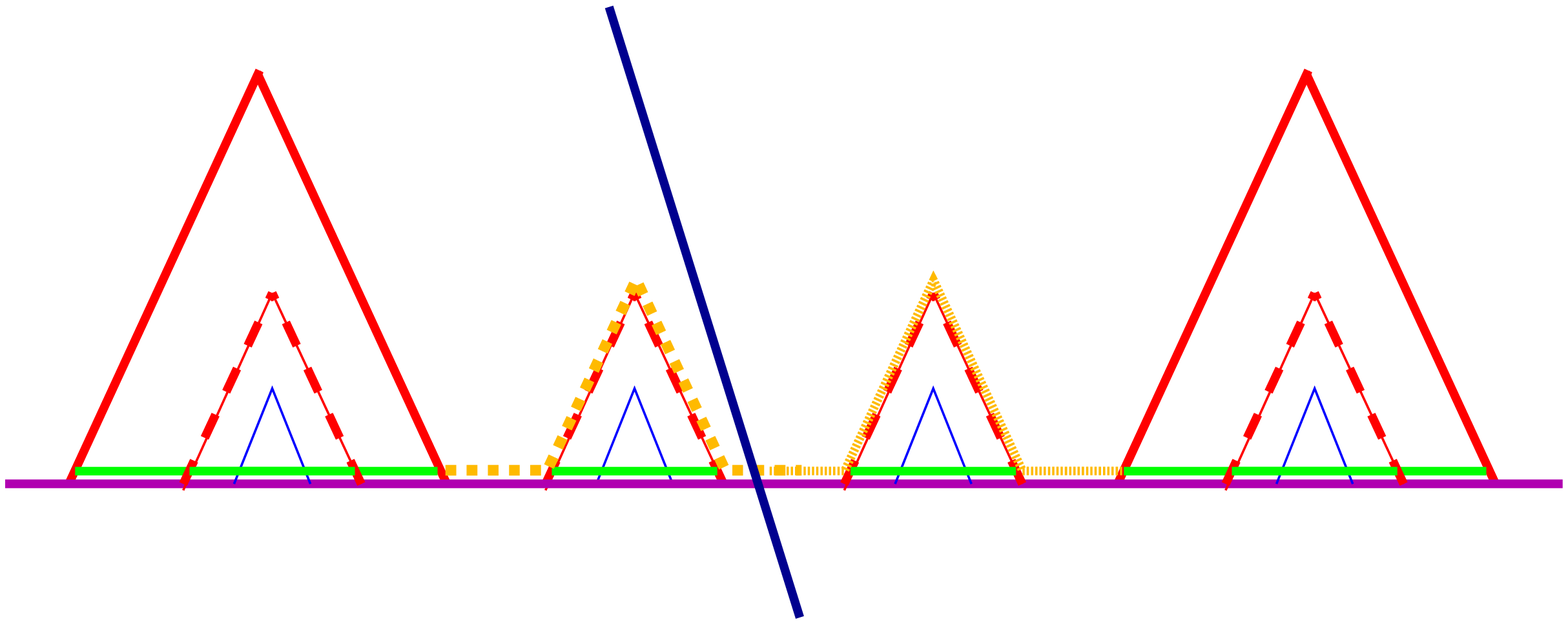,width=300pt}}
\ligne{\hfill
\PlacerEn {-130pt} {0pt} \box110
\hfill}
\vspace{-15pt}
\begin{fig}\label{antennas}
\leurre
The principle of the antennas.
\end{fig}

   Now, a triangle of the next
generation would be raised in the mid-line of a triangle~$K$ 
of~$\Lambda$ and inside~$K$. This is impossible as there is no
triangle of~$\Lambda$ between~$A$ and~$B$. And for still bigger
trilaterals, if both legs are crossed, the trilateral contains
a triangle of the generation of~$T$ within~$\Lambda$ too. Indeed,
once a trilateral exists, it contains all possible trilaterals
which can be contained within its area. 

Accordingly, when $I$ crosses both legs of a trilateral, it is 
a phantom and, more over, the mid-line of this phantom is supported
by~$I$. This is a consequence of the properties of silent
intervals.

   From this analysis, we conclude that there is a tile~$C$
on~$I$ between~$A$ and~$B$ such that on the left-hand side 
of~$C$ with the following property. Considering the
crossing of~$I$, if any, by legs of trilaterals of bigger 
generation, we have that the intersected right-hand side legs 
are all on the left-hand side of~$C$ and that the intersected
left-hand side legs are all on the right-hand side of~$C$.
In fact there are several such tiles~$C$ which constitute
an interval of~$I$, between~$A$ and~$B$. Let us denote
this interval by~$[C,D]$.\cqfd

   As phantoms sharing the same mid-line are constituted
in towers, within a latitude~$\Lambda$, a tower is necessarily finite
and so, it contains an {\bf eldest} phantom.

   The antenna coming from~$T$ will cross the right-hand side
legs which are on the left-hand side of~$C$ and it will jump 
over all eldest phantoms whose both legs are crossed by~$I$ and 
which stand between~$A$ and~$C$. The antenna coming 
from~$T'$ will do the same for the eldest phantoms whose both
legs are crossed by~$I$ and which stand between~$B$ and~$D$.
In between $C$ and~$D$, there are only phantoms within~$\Lambda$.
The antennas go on jumping other the eldest ones they meet
until both antennas meet on some tile of~$[C,D]$, outside
any phantom, see figure~\ref{antennas}.

\vskip 5pt
{\leftskip 20pt\parindent 0pt\baselineskip 11pt
{\bf Remark}. We can prove that in between two contiguous triangles
of the same generation and within the same latitude~$\Lambda$, 
there is at most one leg of a higher generation crossing~$A$ and~$B$:
either of the generation~$n$+1 or of the generation~$n$+2. It
belongs to a generation generated by one of the two triangles
or generated by the former one.

\parindent 20pt
   This can be proved as follows. 

   First, say that a {\bf triangle}~$T$ is the {\bf father} of
a {\bf trilateral}~$K$ if $T$ is of the generation~$n$ and $K$~is of
the generation~$n$+1 and if the vertex of~$K$ is on the mid-line 
of~$T$ inside~$T$. Note that a trilateral has a unique father 
and that a triangle has a lot of sons. We can repeatedly apply this
definition to a trilateral~$K$, leading to a sequence $T_0,\dots T_m$
of triangles with $T_0$ of the generation~0, such that $T_i$~is 
the father of~$T_{i+1}$, for $i\in\{0..m$$-$$1\}$ and that
$T_m=K$. In this case, $T_0$ is called the {\bf remotest ancestor}
of~$K$.

   From the study performed in \cite{mmnewtechund} and from the
results indicated in sub-ection~1.2 of this paper, the height of 
a trilateral of the generation~$n$ is $2^{n+1}$+1 if measured
by the number of the isoclines crossing its legs, vertex and basis
being included in this account. Now, it is not difficult to show
that the distance of the mid-line of a trilateral to the vertex
of its remotest ancestor is $2^{n+1}$, measured in the same way:
we count the isoclines crossing the legs from the mid-line to
the vertex of the remotest ancestors, the last vertex and the 
initial mid-line being taken into account.

Let $T_1$ and~$T_2$ be the 
contiguous triangles of exactly the same latitude and let~$A$ 
and~$B$ as in lemma~\ref{horizontal}. Assume that a 
trilateral~$P$ whose basis is also generated by~$T_1$ exists
and that the right-hand side leg~$\delta$ of~$P$ crosses $AB$, 
taking into account that, at a corner on~$AB$, we consider that 
the leg crosses~$AB$. Then, the ancestor~$A_1$ of~$T_1$ is on the 
closest isocline~0 to the mid-line~$\lambda$ of~$P$, under~$\lambda$,
and the ancestor~$A_2$ is also on this isocline. Now, $A_2$ is 
outside~$P$, otherwise $T_2$ would also be inside~$P$, which
contradicts the assumption. Now,
there is no other seed on the isocline~0 between the vertices of~$A_1$
and~$A_2$. If there would be another one~$\sigma$, $\sigma$ would
be the vertex of a triangle~$A_3$ of the generation~0 and, the same
trilaterals as those occuring inside the trees rooted at the vertices 
of~$A_1$ and~$A_2$ would also occur inside the tree rooted at the
vertex of~$A_3$. In particular, there would be a triangle~$T_3$
of the generation of~$T_1$ and in the same latitude and which would
stand between~$T_1$ and~$T_2$. This is impossible by our assumption.

   Now, if there would be another leg~$\ell$ crossing $AB$
belonging to a trilateral of the generation~$m$ with $m>n$,
then $\ell$~would be between~$\delta$ and~$A_1$, for instance.
It is easy to see that it must be a right-hand side leg.
Otherwhise, the trilateral~$Q$ defined by~$\ell$ would contain~$A_2$ 
as, were it not be the case, this trilateral would contain a copy 
of~$T_1$ which would stand in between~$T_1$ and~$T_2$, again a 
contradiction. But if $Q$~contains~$A_2$, it also contains
a trilateral generated by~$T_2$, which is a copy~$R$ of~$P$ and,
necessarily, $R$ would stand outside~$P$. Now, whatever the
distance between the roots of~$P$ and~$Q$, the distance at
the mid-points, which are on the same isocline
is much bigger, and there is room for an active seed~$A_3$ 
in between~$A_1$ and~$A_2$:  see, below, the table of distances 
between the border of a tree and the closest outside seed on an
isocline~0: 2, 36 or~269. This would again contradict our assumption
on~$T_1$ and~$T_2$.

And so, we may assume that $\ell$~is the right-hand side leg
of a trilateral of the generation~$n$+2 generated by the mid-line 
of~$P$ and inside~$P$. Is this possible? 

The mid-line of~$P$ is~$\lambda$. Let $\nu$ be its closest active 
seed near the right-hand side leg~$\ell$ of~$P$. The problem which 
we have to consider is the distance between the closest seed 
to~$\ell$. We measure this distance in the number of nodes
from~$\ell$ to the seed which are on the same isocline. In fact,
we have to consider the distance on~$\lambda$ and on the next
isocline~0.

   From the study of \cite{mmnewtechund} performed with the
help of a computer programme, and taking into account the
periodicity of the tiles on the border of a tree of the mantilla,
the considered distances are given by the following table:

\vskip 5pt
\vtop {
\ligne{\hfill\hbox to 50pt{\hfill $(\lambda)$, 15:\hfill}
             \hbox to 50pt{\hfill 2\hfill} 
             \hbox to 50pt{\hfill 36\hfill} 
             \hbox to 50pt{\hfill 269\hfill} 
       \hfill}
\ligne{\hfill\hbox to 50pt{\hfill 0:\hfill}
             \hbox to 50pt{\hfill 36\hfill} 
             \hbox to 50pt{\hfill 269\hfill} 
             \hbox to 50pt{\hfill 2\hfill} 
       \hfill}
}
\vskip 5pt

   Also note that for the isoclines~15, these distances give
the closest seed, which is not necessarily active.
   Accordingly, we can see that, whatever the distance of the 
closest active seed~$\sigma$ to the leg~$\ell$ of the 
generation~$n$+1 on the isocline~15, the closest seed to~$\ell$ 
on the isocline~0 is in between the right-hand side leg of the 
tree rooted at~$\sigma$ and~$\ell$. This means that the remotest 
ancestor~$A_1$ of~$T_1$ is not inside the tree rooted at~$\sigma$. 
And so, the righ-hand side border of this tree is on the left-hand 
side of the left-hand side border of~$T_1$.

   This indicates that another possible leg of a higher generation
between~$T_1$ and~$T_2$ could be of the generation~$n$+2
and, precisely, in such a case, the remotest ancestor of~$T_1$
would be the rightmost side on the isocline~0 inside the tree
rooted at~$\sigma$.

   Now, the distances of the closest seeds on the isocline~15 and
the next isocline~0 outisde a leg of a trilateral are given
by the same table as above. Accordingly, between $\delta$ 
and~$T_1$, there
cannot be a leg of the generation~$n$+2. Higher generations 
are a fortiori ruled out: otherwise, the generation~$n$+2
would also be present.

   A last point to notice is that there cannot be two legs
of the generation~$n$+1: a right-hand side one on the side of~$T_1$
and a left-hand side one on the side of~$T_2$.

Now, the distance between legs of such opposite legs of the same
generation is increasing as we go down along the isoclines.
The closest distance is the smallest distance bewteen two
active isoclines. It is not difficult to see that it is realized
by the $F$-sons of a~$G_r$- and a~$G_\ell$-centres which belong to
the same $F$-flower when the $F$-centre is a black node. The
distance is then~26. At the level of~$\lambda$, the distance is 
much bigger than the biggest distance~269 indicated by the above 
table. 

   Accordingly, two such legs cannot be present between~$T_1$
and~$T_2$.

   And so, our claim is proved.
\par}

   Next, we provide tiles to implement this mechanism and then,
in the next subsection, we check that the new mechanism does
not disturb the general construction, outlined in the previous
section.

\subsection{The tiles}

   The antenna is given a specific colour, we call it {\bf orange},
and it has a laterality: there is an right-hand side antenna, which
goes to the right and a left-hand side one, which goes to the left.
Due to the colour of the antenna, we shall often speak of the
{\bf orange signal} and of its laterality. 

   The first principle is that an antenna cannot directly be in
contact with the green signal. And so the green signal and the
orange one are always separated by the leg of a trilateral, more
precisely, they both occur at the mid-point of a leg. The other
principle is that the antenna is stopped, at one end by a
mid-point of a triangle and, at the opposite end, by
an antenna of its opposite laterality.

\vspace{-15pt}
\setbox110=\hbox{\epsfig{file=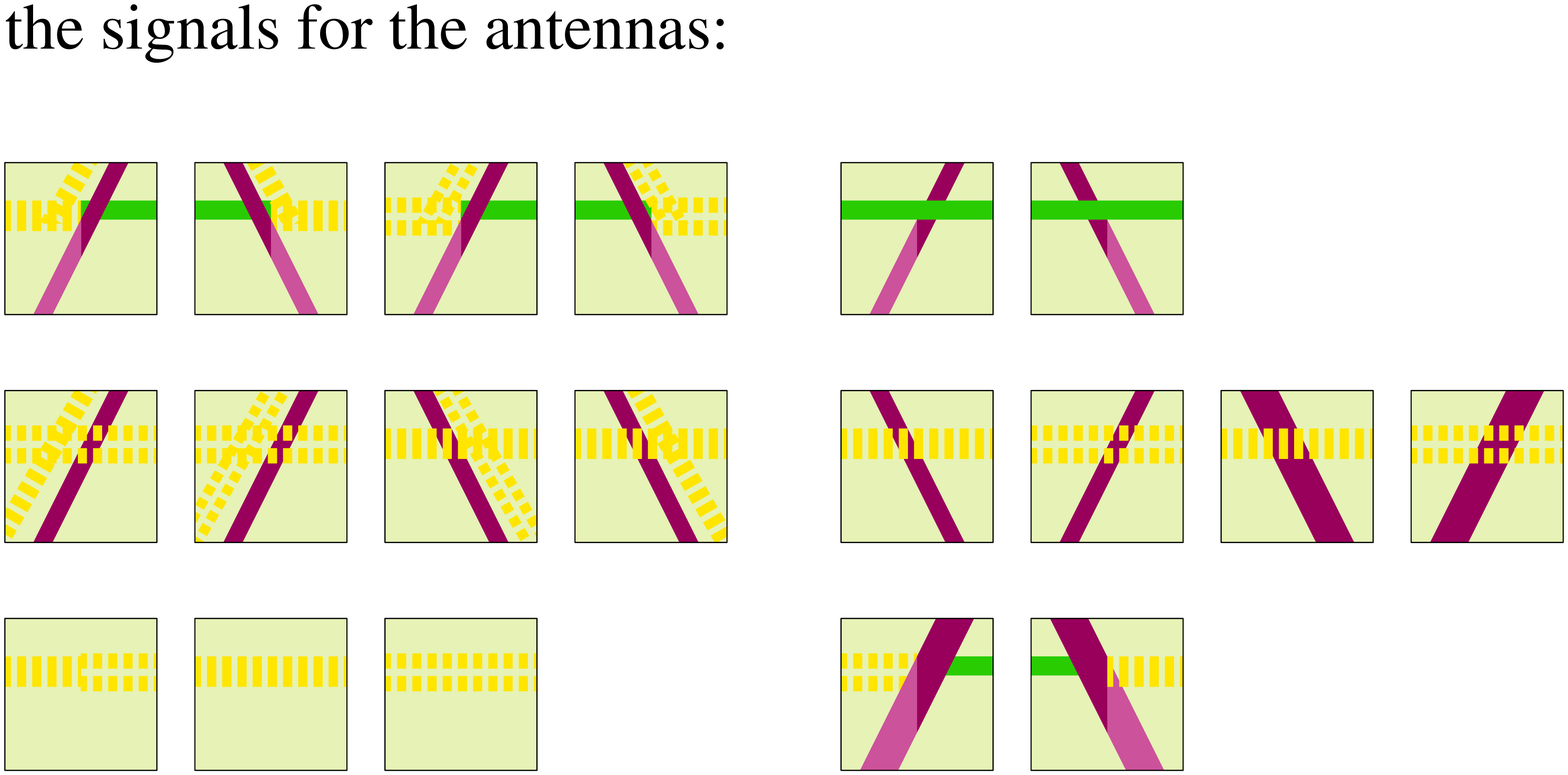,width=300pt}}
\ligne{\hfill
\PlacerEn {-150pt} {0pt} \box110
\hfill}
\vspace{-15pt}
\begin{fig}\label{tile_antennas}
\leurre
The tiles for the antennas:
\vskip 1pt
\noindent
Number them by rows and columns, in $1..3$ for rows and
$1..6$ for columns.
\vskip 0pt
\noindent
In the row~$3$, we have the ends of the antennas: the join tile,
$1-1$ and the mid-point of a triangle, $3-5$ and $3-6$.
\vskip 0pt
\noindent
In the row~$1$, we have the tile to jump over an eldest phantom.
\vskip 0pt
\noindent
In the row~$3$, we have the tiles to cross legs of bigger trilaterals.
Note that not any combination of lateralities for the leg and for the
antenna are permitted. Note that the two tiles for the vertices are
not represented, as this is easy.
\end{fig}
\vspace{-10pt}

   Now, if the antenna meets a leg at a mid-point, and there is
necessarily a green signal on the other side of the leg, then the
orange signal climbs up along the leg. 

   The tiles are dispatched by figure~\ref{tile_antennas}.
 
\subsection{Checking the correctness}

   We have now to check that the set of tiles given by 
figure~\ref{tile_antennas} force the above specifications given to
the antennas. Note that the tiles are in fact {\bf meta-tiles}.
We represented the colour by a variable colour. Note also that
in the tiles of the row~1, we have mid-points of phantoms, 
necessarily. However, in the row~2, tiles 2-5 and 2-6 belong to 
legs of phantoms, while tiles~2-7 and~2-8 belong to legs of triangles.
In both cases, it may be either the first half or the second half
of a leg. The tiles 2-1, 2-2, 2-3 and 2-4 concern phantoms only and,
as the other tiles of the row, of any colour and for
any half-leg. Also, the two meta-tiles for the vertices
are not represented. One is endowed with the signal to the right, the
other with the signal to the right. Note that a vertex cannot join
antennas of different lateralities. Note that the vertex
of a phantom with no orange signal over it is also available. 

   It is not difficult to check that the antennas can be constructed
by the considered tiles. We shall focus on the converse: nothing else
can be obtained from the tiles.

   First, we check that inside an eldest phantom, the phantoms are 
crossed at their mid-line by the green signal only. This is
obtained by the combination of lateralities and the fact that
both first halves of the legs of a phantom are covered by a signal
with the same laterality. Also note that there is a single join-tile
for horizontal parts of antennas of opposite laterality. This
join-tile prevents to change of laterality inside a phantom.
Accordingly, if an orange signal would cover a non eldest phantom, 
there would be a contradiction at one mid-point of a the phantom of 
the next generation: a contradiction on the left-hand side mid-point
with an orange signal to the right and on the right-hand side 
mid-point with an orange signal to the left. 

   The same argument
explains that the present construction does not prevent the
mechanism of the green signal to detect the mid-line of a triangle.
Indeed, if instead of the tile~3-5 a tile 2-8 is used, as the
join tile 3-1 cannot be used inside the triangle, the orange signal,
after jumping other the eldest phantoms inside the triangle 
could not match the meeting with the other leg of the triangle.
We have a symmetrical argument if the tile 2-7 is used in place
of the tile~3-6. And so, the unique solution is to use the tiles~3-5
and~3-6.

   Also, as the orange signal cannot directly meet with a green one,
on a given mid-line, we have either the green signal or an 
 orange one. We have a green signal inside a triangle and inside
the phantoms whose mid-line is the considered isocline.

   We also note that the tiles of the row~2 cannot be used in place
of those of the row~1 and that the tiles of the row~2 must be used
with legs of trilaterals of a bigger generation than that of the
triangles emitting the crossing signal. Note that the tiles 2-1,
2-2, 2-3 and 2-4 are used by triangles contained in an eldest 
phantom~$P$.  As the mid-lines of these triangles is not the mid-line
of~$P$, the signal crosses the leg which bears the orange signal
covering~$P$. Note that the opposite legs of~$P$ are crossed by
signals of the same laterality as the leg and that due to the
unique join tile, there must be a triangle inside the considered
phantom. Also note that for the phantoms of the generation~0 this
brings in no contradiction as they do not contain triangles. 

   At last, the start of a jump at a mid-point cannot be confused
by a crossing. For instance, if the antenna to the righ from~$T$
goes to far and meets a left-hand side leg on the right-hand side 
of the point~$D$ defined at sub-section~2.2, then there is a 
trilateral which receives a green signal which will meet the orange 
signal of the left-hand side antenna from~$T'$, which will produce
a contradiction. and so, the single solution is to use the joining
tile at a place of~$[C,D]$ which is outside any phantom of the
considered latitude.

   And so, this proves that the antenna mechanism is forced by the
set of tiles of figure~\ref{tile_antennas}.

   Note that in the case of the butterfly model, see 
\cite{mmarXiv2,mmnewtechund}, the mechanism of the antenna forces 
the green signal to run over the whole isocline
which is the mid-point of the latitude which contains no triangle.
Indeed, the laterality constraints of the tiles of the second row
prevent an orange signal to run at infinity.

   With this, we completed the proof of theorem~\ref{undec}.

\section*{Conclusion}

   The first consequence is that we need a bigger number of tiles
than what is announced in \cite{mmarXiv2,mmnewtechund}. Indeed,
the orange signals entails an increase of the number of vertices,
of mid-points, of corners, of crossings with various legs. Signals 
of bases are also changed by the completion of the construction. 
Note that corners behave as crossing legs. Also note that
the signal particularly addresses isoclines~5 and mainly~15.
A detailed counting will be given in a forthcoming paper making the
synthesis of \cite{mmarXiv2,mmnewtechund} and the present paper.
However, a rough estimate shows that the number of prototiles
should be now around 21,000 tiles. But, the number of meta-tiles,
the variable tiles indicating the computation signs which depend
on the simulated Turing machine, is not changed by the orange
signal.

   It is interesting to notice the r\^ole played by the laterality
in the whole proof of theorem~\ref{undec}. The laterality
is not used exactly in the same way in the antenna mechanism and
in the mechanism of detecting the green line, the bases and 
the free rows. However, the same difference is used together
with the possibility to connect opposite lateralities in a single 
way. May be a closer analysis of this mechanism could be used
to reduce the number of signals, hence to reduce the number of
tiles.

   In this line, it is certainly possible to change a bit
the construction of the triangles of generation~0. There is no need to 
use a green signal, the meeting with an isocline~5 is enough to
play this r\^ole. The advantage is that the occurrence of
the green and orange signals would be restricted to the
isoclines~15. accordingly, the scent would trigger the green
signal only by meeting an isocline~15.

\def\kvs{\vspace{-3pt}}

\end{document}